\documentclass[aps,prapplied,reprint,superscriptaddress,twocolumn]{revtex4-2}
\usepackage[utf8]{inputenc}
\usepackage[T1]{fontenc}
\usepackage{textcomp, mathtools, amssymb, amsthm, wasysym, gensymb}
\usepackage[colorlinks=true, linkcolor=red, citecolor=blue, urlcolor=blue]{hyperref}

\newcommand\unine{Laboratoire Temps-Fréquence, Institut de Physique, Université de Neuchâtel, Avenue de Bellevaux 51, 2000 Neuchâtel, Switzerland}
\newcommand\ethz{Institute for Quantum Electronics, ETH Zurich, Auguste-Piccard-Hof 1, 8093 Zurich, Switzerland}
\newcommand\nanoqt{Present address: Nanofiber Quantum Technologies, Inc., 1-22-3 Nishiwaseda, Shinjuku-ku, 169-0051 Tokyo, Japan}

\begin{document}

\title{Non-resonant Optical Injection Locking in Quantum Cascade Laser Frequency Combs}

\affiliation{\unine}
\affiliation{\ethz}
\affiliation{\nanoqt}

\author{Alexandre~Parriaux}
\email[Corresponding author: ]{alexandre.parriaux@unine.ch}
\affiliation{\unine}

\author{Kenichi~N.~Komagata}
\affiliation{\unine}
\affiliation{\nanoqt}

\author{Mathieu~Bertrand}
\affiliation{\ethz}

\author{Mattias~Beck}
\affiliation{\ethz}

\author{Valentin~J.~Wittwer}
\affiliation{\unine}

\author{Jérôme~Faist}
\affiliation{\ethz}

\author{Thomas~Südmeyer}
\affiliation{\unine}

\date{\today}

\begin{abstract}
Optical injection locking of the repetition frequency of a quantum cascade laser frequency comb is demonstrated using an intensity modulated near-infrared light at 1.55~µm illuminating the front facet of the laser. Compared to the traditional electrical modulation approach, the introduced technique presents benefits from several perspectives such as the availability of mature and high bandwidth equipment in the near-infrared, circumvent the need of dedicated electronic components for the quantum cascade laser, and allows a direct link between the near and mid-infrared for amplitude to frequency modulation.
We show that this stabilization scheme, used with moderate near-infrared power of a few milliwatts, allows for a strong reduction of the frequency noise. We also perform a full characterization of the mechanism and evidence that the locking range follows Adler's law. 
A comparison of our results with those in recent literature indicates that the optical approach leads to better performance compared to the traditional method, which we expect to benefit mid-infrared spectroscopy and metrological applications. 
\end{abstract}

\maketitle

\section{Introduction}
Frequency combs, i.e, coherent sets of equally spaced discrete optical frequency lines, can be generated with a variety of sources~\cite{Fortier-commphys-2019}, such as mode-locked lasers~\cite{Kim-aop-2016}, micro-resonators~\cite{Pasquazi-physrep-2018} or electro-optic modulators~\cite{Parriaux-aop-2020}. In the mid-infrared (MIR), combs sources are scarcer and this spectral region is usually reached via nonlinear frequency conversion from the near-infrared (NIR)~\cite{Sotor-oe-2018,Chen-pnsa-2019,Hoghooghi-lsa-2022}. However, a few candidates like quantum cascade lasers (QCL) are able to generate a frequency comb directly in the MIR~\cite{Hugi-nature-2012,Faist-nanophot-2016}, and these particular lasers present unique properties such as a low footprint, high repetition frequencies, watt-level output power~\cite{Jouy-apl-2017}, potential for mass production, etc. This makes them particularly interesting in several applications such as high resolution spectroscopy~\cite{komagata2022absolute,hayden-apl-2024}, time resolved spectroscopy~\cite{klocke-analchem-2018}, free-space communications~\cite{Corrias-oe-2022} or biomedical spectroscopy~\cite{Schwaighofer-chemsocrev-2017}.

Nonetheless, as with other lasers, QCLs are subject to noise types of several origins which can be linked to their structure~\cite{Schilt-apb-2015}, and the temperature of operation and the related internal fluctuations~\cite{Bartalini-prl-2010,Bartalini-oe-2011,Borri-jqel-2011,tombez-oe-2012,tombez-apl-2013}, to only cite a few examples.
This degrades the high spectral purity of single-frequency QCLs and limits their use for most applications as mentioned above.
Combs generated by QCLs (QCL-comb) suffer from similar problems~\cite{Hugi-nature-2012,Cappelli-optica-2015,Shehzad-oe-2020}, which means that to improve performance, stabilization procedures are required on both degrees of freedom of the comb, namely the repetition frequency $f_\text{rep}$ and the offset frequency $f_0$. For a MIR QCL-comb, the latter is technically difficult to detect directly via standard $f-2f$ interferometry~\cite{jones-science-2000,reichert-prl-2000} due to low peak power generated by QCLs that prevents nonlinear broadening.
However, several scheme addressed this issue by bringing a fully stabilized comb in the spectral region of the QCL-comb allowing its full referencing~\cite{Consolino-natcomm-2019,Chomet-optica-2024}, or by using a MIR molecular reference~\cite{komagata2022absolute}.

As for the stabilization of $f_\text{rep}$, several possibilities are available such as actuation on the current driving the laser in a phase lock loop~\cite{Shehzad-oe-2020}, electrical radio-frequency injection locking (RFIL) \cite{Stjean-lpr-2014,Hillbrand-natphot-2019,Schneider-lpr-2021} or actuation using off-resonant light~\cite{Consolino-lpr-2021}.
Note that recent works demonstrated passive stabilization of the repetition frequency using optical feedback~\cite{teng-ol-2023,huang-oe-2024}, however this approach does not reference $f_\text{rep}$ to a radio-frequency (RF) standard.
Although the first possibility (drive current actuation) is the most straightforward, it was demonstrated that phase locking the repetition frequency to a RF standard induces an increase in the noise of the comb lines~\cite{Shehzad-oe-2020}.
Regarding electrical RFIL, this technique requires dedicated electrical components for the QCL, and we also recently showed limits in the faithful reproduction of the master oscillator through electrical RFIL~\cite{Parriaux-apr-2024}.
This calls for exploring and carefully characterizing different stabilization methods. In our previous study~\cite{Komagata-aplphot-2023}, we showed that NIR light illumination could be one of these new possibilities.

Here, we present the optical RFIL of the repetition frequency in a QCL-comb using a non-resonant, intensity modulated NIR laser illuminating the front facet of the QCL. We characterize the efficiency of the approach in terms of locking range and spectral purity, as a function of injected power. In parallel we also investigate several properties of the QCL-comb under NIR light illumination to fully characterize the impact of its use.

\section{Experimental setup}

\begin{figure}[ht]
\centering
\includegraphics[width=\linewidth]{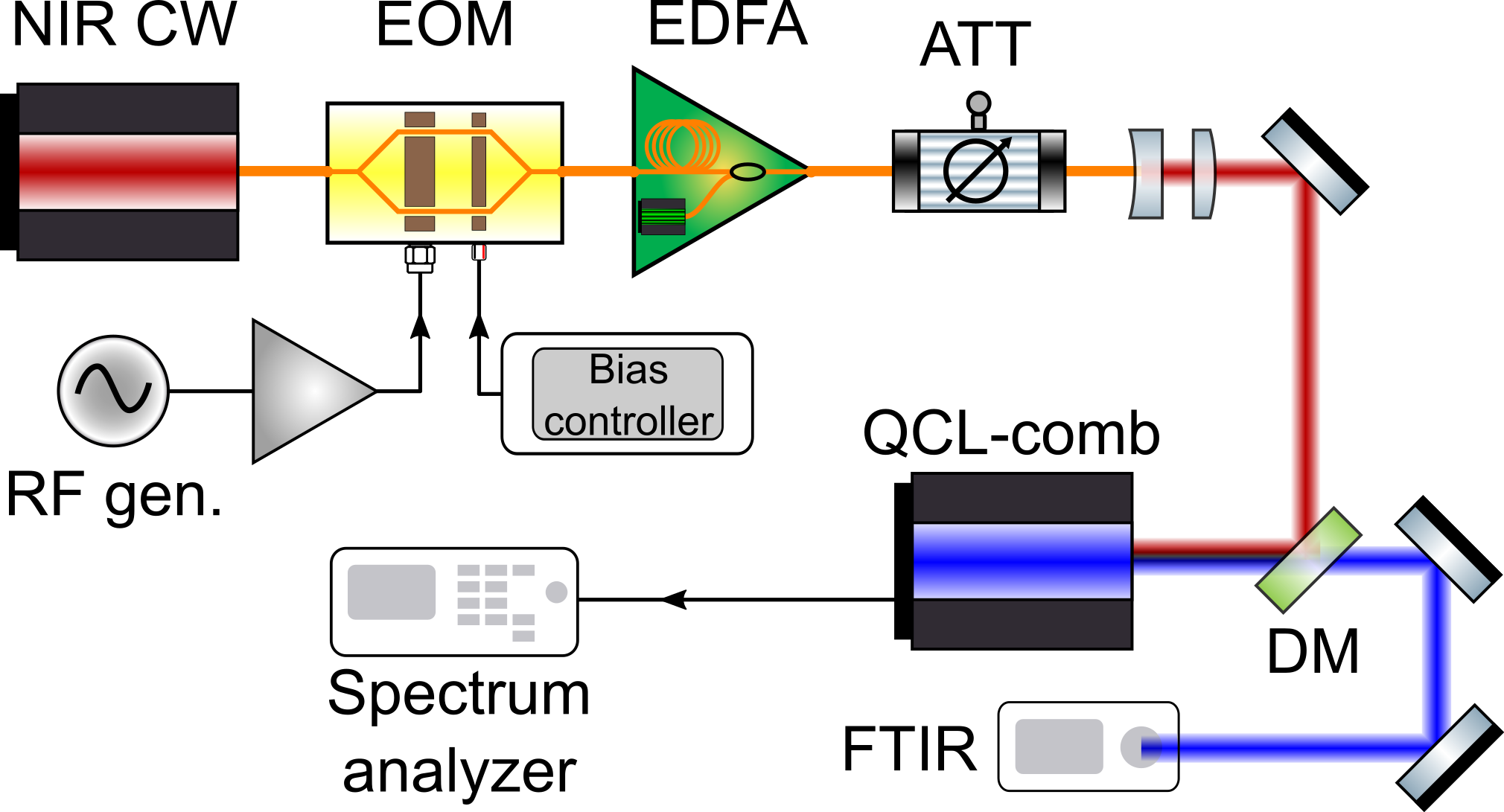}
\caption{Schematic showing the experimental setup used for optical RFIL of the repetition frequency of the QCL-comb using an intensity modulated NIR light illuminating the front facet of the QCL, and for characterizing the comb properties. ATT: variable attenuator, DM: dichroic mirror.}
\label{fig:setup}
\end{figure}

Our experimental setup is divided into two parts and schematized in Figure~\ref{fig:setup}. The first one starts with the generation of a MIR frequency comb centered around 1323~cm$^{-1}$ with a 45~cm$^{-1}$ bandwidth using a QCL driven at 1020~mA and temperature stabilized at 2\textdegree C. The QCL-comb beam is then collimated with a 1.868~mm focal-length aspheric lens and transmitted through a home-made 1.55~µm/7.8~µm dichroic mirror produced with an ion beam sputtering machine. The spectrum of the QCL-comb is monitored using a Fourier-transform infrared spectrometer (FTIR, Bristol 771), and its optical power with a thermal sensor (Thorlabs, S401C).
The repetition frequency of the QCL-comb $f_\text{rep}$ is measured thanks to wires placed on top of the QCL-chip that allows to capture the associated voltage originating from the modes beating in the Fabry-Perot cavity~\cite{rosch-apl-2016,piccardo-optica-2018}. The signal of the electrically extracted beatnote is then characterized using a phase noise analyzer (Rohde \& Schwarz, FSWP26) that allows us to analyze the RF spectrum and the phase noise power spectral density of $f_\text{rep}$.
With this, one can study the performance of the locking scheme that will be introduced shortly after. 

The second part of our setup consists of an intensity modulated and amplified NIR laser that illuminates the front facet of the QCL. Compared to our previous studies~\cite{Komagata-aplphot-2023,Parriaux-apr-2024}, the characteristics of the NIR light are adapted to perform optical RFIL of the QCL-comb repetition frequency $f_\text{rep}$. For this, we use a typical telecommunication laser diode emitting a continuous wave (CW) around 1.55~µm with 40~mW of optical power. The CW laser is then intensity modulated with a 20~GHz electrical bandwidth electro-optic modulator (EOM) (iXblue, MXAN-LN-20) electrically driven by a signal generator at a frequency $f_\text{gen}$ (Rohde \& Schwarz, FSWP26). The bias voltage of the EOM is set at its quadrature point and stabilized using a commercially available bias controller (iXblue, MBC-AN-LAB-A1). 
The RF power used to drive the EOM is set at 20~dBm which maximizes the power of the RF signal seen by a photodiode when detecting the optical output of the EOM. This value of RF power will be kept identical throughout the studies presented here.
Then, the modulated optical signal is amplified up to 70~mW using a home-made erbium doped fibered amplifier (EDFA), and the resulting signal is connected to a variable attenuator to control the optical power that will further illuminate the front facet of the QCL. Note that NIR powers measured and presented in this article are average powers.
The NIR light is then sent to free space via a fibered collimator, and before reaching the home-made 1.55~µm/7.8~µm dichroic mirror introduced previously, we focus the beam using a 300~mm focal-length lens. The dichroic mirror is used in reflection for the NIR beam which is then directed to illuminate the front facet of the QCL with a beam diameter slightly below 20~µm.
Note that a smaller beam size could be obtained with different optics to increase the NIR intensity delivered on the QCL front facet.

\section{Injection locking characterization}

\subsection{Response of the QCL with high power illumination}

\begin{figure}[t]
\centering
\includegraphics[width=\linewidth]{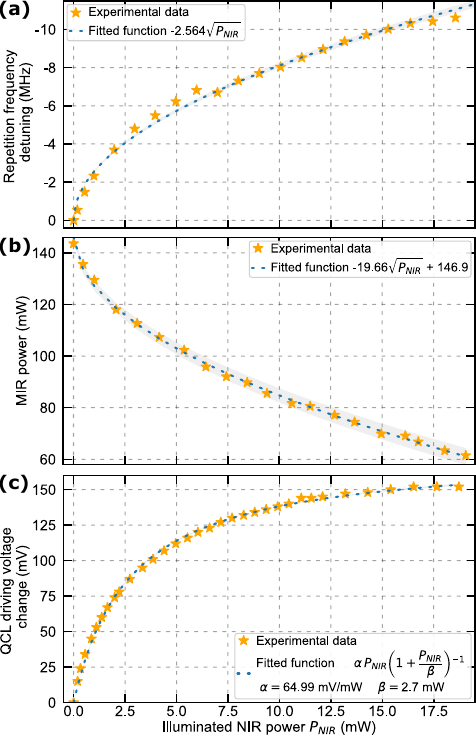}
\caption{Graph showing several parameters of the QCL-comb when illuminated with NIR light, and comparisons with fitted functions.
\textbf{(a)} Repetition frequency detuning. Here the zero frequency detuning corresponds to 11.0601~GHz.
\textbf{(b)} MIR optical power delivered by the QCL.
\textbf{(c)} Change in voltage measured at the bounds of the QCL. The zero voltage change corresponds to a driving voltage of 11.639~V and the driving current was kept at 1020~mA. For all sub-figures, the gray shaded areas represent the two standard deviation margin on the parameters estimated numerically.}
\label{fig:param_qcl}
\end{figure}

In our previous article~\cite{Komagata-aplphot-2023}, we studied the response of several parameters of the QCL when a low power NIR light illuminated its front facet and we showed that the NIR beam alignment was important. Here, as we take an interest in the stabilization of $f_\text{rep}$, the alignment of the NIR beam on the QCL front facet is made so as to maximize the frequency detuning of $f_\text{rep}$ induced by the NIR illumination. We observed that this is directly linked to the maximization of the electrical voltage driving the QCL and several other features (see below).
With such an alignment, the variation of $f_\text{rep}$ with respect to the illuminated NIR power $P_\text{NIR}$ can be studied and is presented in Figure~\ref{fig:param_qcl}~\textbf{(a)}. A comparison of the experimental data with a fitted function shows that the frequency detuning induced by the NIR light closely follows a square root trend of the NIR power. Note that the parameters of the fitted function here and in the following are obtained using a nonlinear least squares regression method to fit the function to the experimental data. Moreover, the standard deviation margins (grey shaded areas in all the figures of this article) are calculated from the covariance matrix resulting from the least squares regression analysis.

In a similar way, and using a thermal sensor, we monitored the optical power emitted by the QCL when illuminated with our NIR light. The MIR power evolution with respect to $ P_\text{NIR} $ is shown in Figure~\ref{fig:param_qcl}~\textbf{(b)}, and a comparison of the experimental data with a fitted function shows a square root function trend as well. As for the repetition frequency detuning, we also observed here that the alignment of the NIR beam on the front facet of the QCL is important, and that the MIR power at a given NIR power is minimized when the driving voltage of the QCL is maximized by NIR light illumination. The trend predicts a complete shutdown of the laser at 60~mW.
Since our QCL operates at a fixed current and well above threshold, the MIR output power losses observed with respect to NIR illumination is due to the changes induced in the density of carriers in the active region of the QCL~\cite{Suchalkin-apl-2013}. Intensity modulation of QCLs using NIR light front facet illumination is not something new and has already been investigated several times~\cite{Chen-apl-2009,Yang-ol-2013}.

Since the evolution of the QCL repetition frequency $f_\text{rep}$ and optical power with respect to $ P_\text{NIR} $ are related to changes in the voltage driving the QCL (at a fixed current), we also studied its evolution in the same NIR beam alignment configuration as above. This is shown in Figure~\ref{fig:param_qcl}~\textbf{(c)} and data analysis shows that the induced voltage closely follows a similar trend as in equation~\eqref{eq:vinduce}. We also observe above 17~mW a plateau which might indicate, at least locally, a saturation effect. Complementary experiments at different working points and using different QCL-chips would be necessary to assess the full picture about saturation.

The behavior of the QCL and the trends observed above when illuminated with high power NIR light is not simple to interpret without performing numerical simulations that investigate the interaction between the QCL and the NIR light, which is beyond the scope of this paper. However, we note that the trends put into evidence here for $f_\text{rep}$, the MIR optical power, and the driving voltage were also observed for other QCL-comb chips, which corroborates the general applications of our findings.
We also note that the evolution of $f_\text{rep}$ and optical power shows similar square root trends as observed for the optical central frequency evolution of terahertz CW QCLs illuminated with a NIR light as shown in Ref.~\cite{alam-oe-2019a,alam-oe-2019b}.

Finally, we can explain the behavior observed in Figure~\ref{fig:param_qcl}~\textbf{(c)}, which will be useful to understand the features of the locking mechanism under study here. When illuminating the front facet of the QCL, part of the NIR light is absorbed in the active region and modulates the density of carriers. To be more precise, the available carriers in the quantum wells are depleted and at some point, the absorption saturates~\cite{fox-jqel-1991,sizer-jqel-1994}. Following the approach used in Refs.~\cite{fox-jqel-1991,sizer-jqel-1994,kruger-oe-2021}, we can define an absorption coefficient per unit length $\alpha$ as:
\begin{equation} \label{eq:coeffabs}
    \alpha(P_\text{in}) = \alpha_0 \left ( 1 + \frac{P_\text{in}}{P_\text{sat}}  \right )^{-1}
\end{equation}
where $P_\text{in}$ is the NIR power effectively seen by the active region, $\alpha_0$ the absorption coefficient at low power, and $P_\text{sat}$ the saturation power at which $\alpha_0$ is reduced by two. Since we illuminate the front facet of the QCL using a CW light, the photovoltage generated $ V_\text{DC} $ has only a DC component and can be written as:
\begin{equation} \label{eq:vinduce}
    V_\text{DC}(P_\text{in}) = \gamma_{0,\text{DC}} \, P_\text{in} \left ( 1 + \frac{P_\text{in}}{P_\text{sat,DC}}  \right )^{-1}
\end{equation}
where $ \gamma_{0,\text{DC}} $ is the DC conversion gain at low power, and $P_\text{sat,DC}$ the saturation power at which $\gamma_{0,\text{DC}}$ is reduced by two. As can be seen in Figure~\ref{fig:param_qcl}~\textbf{(c)}, the experimental data closely follow the trend defined by equation~\eqref{eq:vinduce}.

\subsection{Frequency noise evolution}
Using the EOM, we now modulate the NIR CW laser for injection locking of the repetition frequency. First, we clarify the physical origin behind the locking mechanism. As previously mentioned, when we illuminate the QCL with NIR light, the density of carriers is modulated in the active region which produces an electrical response driven by the NIR light properties. 
Note that in our case, this modulation is localized in a small region close to the front facet. Consequently, when the NIR light is modulated, a RF signal is generated which, in the end, also results in a modulation of the laser gain in this localized region. At this point, similarly as electrical RFIL which also modulates the laser gain, the natural repetition frequency of the QCL can be locked on the RF signal generator under the right conditions, namely frequency matching and power level, as described by Adler's formalism of coupled oscillators~\cite{Adler-ieee-1946,Siegman-lasers-1986}.

\begin{figure}
\centering
\includegraphics[width=\linewidth]{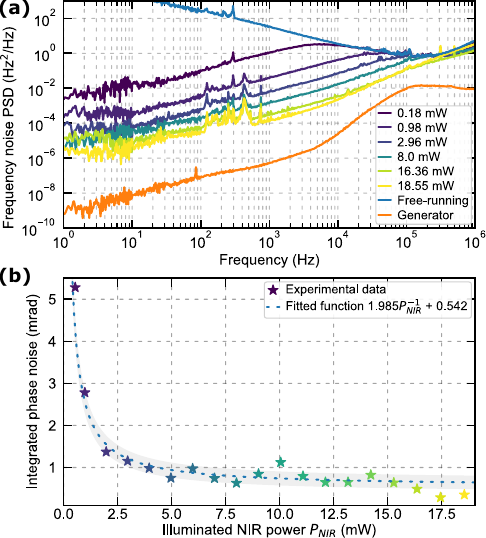}
\caption{\textbf{(a)} FN-PSD of the QCL-comb repetition frequency when injection locked using a modulated NIR light illuminating its front facet. \textbf{(b)}~Integrated phase noise from 1~Hz to 1~MHz related to the different phase noise measurements performed versus the NIR illuminated power $P_\text{NIR}$. The gray shaded area in \textbf{(b)} represents the two standard deviation margin on the parameters estimated numerically.}
\label{fig:phasenoise}
\end{figure}

Our procedure to study injection locking of the repetition frequency is the following. First, using the variable attenuator at the end of the NIR fibered path, we adapt $P_\text{NIR}$ to the desired value. As shown in Figure~\ref{fig:param_qcl}~\textbf{(a)}, the natural repetition frequency of the QCL will be shifted and so we set the frequency of the generator $f_\text{gen}$ to this new frequency. 
When the modulation is turned on, using the phase noise analyzer, we study the RF spectrum of $f_\text{rep}$ that is electrically extracted on the dedicated channel of the QCL, and the associated frequency noise power spectral density (FN-PSD) which is obtained by converting the phase noise measurements. Figure~\ref{fig:phasenoise}~\textbf{(a)} presents different FN-PSD of $f_\text{rep}$ at several illuminated NIR power levels. Note that for clarity and readability of the graph, only a few curves are displayed but we performed measurements at the same points as the ones in Figure~\ref{fig:param_qcl}~\textbf{(a)} (experimental data associated to the non-displayed curves are available in the link provided in the Data Availability section). Also note that we measured the FN-PSD of the RF beatnote delivered by our NIR setup using a fast photodiode and we observed no difference with the FN-PSD of the signal delivered by the generator used to drive the EOM.

In Figure~\ref{fig:phasenoise}~\textbf{(a)}, we observe that even with a power as low as 180~µW, injection locking can be performed and reduces the FN-PSD drastically, especially at low frequencies where it is reduced by 80~dB at 1~Hz. When increasing $ P_\text{NIR} $ the noise at low frequencies continues to decrease and the lock bandwidth, which is the highest frequency at which the noise is reduced compared to the free running state, increases.
To gain more insight, we calculated the integrated phase noise $\phi_\text{int}$ from 1~Hz to 1~MHz associated to the different phase noise measurements performed and plotted it against the NIR illumination power, which is shown in Figure~\ref{fig:phasenoise}~\textbf{(b)}.
A comparison of the experimental data with a fitted function shows that $\phi_\text{int}$ decreases following an inverse function trend with respect to $ P_\text{NIR} $. Hence, higher illumination power is not linked to a large noise reduction, which motivated our choice to stop power studies around 19~mW of illuminated power. Moreover, one should note that even if $ P_\text{NIR} $ can be increased, at a certain point our QCL-comb often started to jump to a different mode, which is associated to radical changes in the optical spectrum and $f_\text{rep}$ and hence detrimental for coherent analysis of the results.
A last remark is that the oscillatory feature observed in Figure~\ref{fig:phasenoise}~\textbf{(b)} is due to the noise contribution between 1~Hz and 10~Hz which impacts the most on the integrated phase noise value, and that can easily vary in minutes long measurements due to mechanical vibrations. Better packaging and thermal regulation of the QCL could reduce and even suppress this effect.

Another interesting feature with this injection locking technique is that the frequency of the generator driving the EOM can be set to half or twice the natural repetition frequency of the QCL-comb. In both cases, we observed only small changes in the FN-PSD of the locked $f_\text{rep}$ compared to the case where the frequency of the generator matches $f_\text{rep}$. We believe that these small variations originate from the non-constant electrical response of our setup over the whole range $\frac{1}{2}f_\text{rep}$ to $2 f_\text{rep}$ which is more than 16~GHz.
Note that this can easily be compensated by a small adjustment of $ P_\text{NIR} $. Although no improvement are observed, this feature can still be useful for designing a cost effective setup as the frequency bandwidth required for the whole NIR setup can be reduced by a factor of 2. Exploring different multiplication or division factors is an interesting topic, but it would deserve a dedicated study.

To better apprehend the results and the impact of using optical RFIL, we can compare the performance obtained with the electrical RFIL approach. This is, however, a difficult task as very little works reported clear FN-PSD measurements of the repetition frequency of a QCL-comb under electrical RFIL for the reasons we explained in Ref.~\cite{Parriaux-apr-2024}. Still, some works evidenced limits of electrical RFIL which we can hereby address for a comparison with optical RFIL, especially with our previous study as we used the same QCL.
These limits, observed in Refs.~\cite{Consolino-natcomm-2019,Parriaux-apr-2024}, first concerns a bump above 1~kHz in the noise measurements of different comb lines under investigation. As stated in these two works, the bump is related to the comb mode spacing properties as it evolves with the line number. In the case of optical RFIL, the results displayed in Figure~\ref{fig:phasenoise}~\textbf{(a)} do not show such a feature, which seems to indicate that optical RFIL has a better locking bandwidth property.
Second, as suggested in Ref.~\cite{Parriaux-apr-2024}, the reproduction of the noise characteristics of the injected oscillator with electrical RFIL seems to be limited to a few kHz offset frequency and with an inverse frequency trend. In the case of optical RFIL, we also observe a similar trend but the frequency limit seems to be at least one order of magnitude higher as can be seen in Figure~\ref{fig:phasenoise}~\textbf{(a)} where we observe no more enhancement of the FN-PSD above 50~kHz when increasing the NIR power.

Finally, to the best of our knowledge, there is only one work that presented FN-PSD measurements of the repetition frequency of a QCL-comb under electrical RFIL using an optical approach to avoid any parasitic effect (see Ref.~\cite{Parriaux-apr-2024}). These measurements are presented in Ref.~\cite{Chomet-optica-2024} and more precisely in the supplementary materials, Fig.~S7~(b)~\cite{Chomet-optica_sup-2024}. By comparing the FN-PSD measurement at the highest injected RF power of 18~dBm with our results, we observe that the FN-PSD measurement obtained using only 1~mW of NIR power already shows enhanced performance.

Hence, the presented results suggest that optical RFIL is superior than electrical RFIL in terms of achievable noise reduction. One of the potential explanation is that compared to electrical RFIL, optical RFIL is less affected by any electrical impedance mismatch when injecting the RF signal. However, we remain cautious with this claim since the accurate measurement of residual phase noise in the case of electrical RFIL has only been recently demonstrated and not optimized over a broad range of parameters as with the current study. Further studies such as the analysis of the electric field emitted by a QCL stabilized with electrical RFIL or optical RFIL could settle down this issue~\cite{Chomet-optica-2024}.

\subsection{Locking range}
\begin{figure}[th]
\centering
\includegraphics[width=\linewidth]{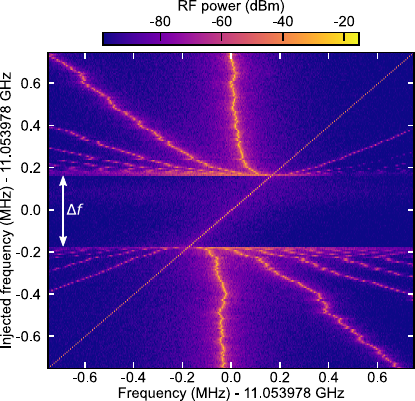}
\caption{Set of RF spectra obtained with a resolution bandwidth of 1~kHz at the electrical extraction channel of the QCL when sweeping the frequency of the generator (diagonal line) to observe the bandwidth $\Delta f$ where the repetition frequency locking occurs. The NIR light power illuminating the front facet of the QCL was set to 6~mW.}
\label{fig:capture_range}
\end{figure}

One last feature that we studied regarding optical RFIL of $f_\text{rep}$ is the locking range with respect to the NIR illumination power, which is defined as the frequency range $\Delta f$ where injection locking is possible. For this, we linearly swept the frequency of the RF generator $f_\text{gen}$ used to modulate the NIR light around the natural $f_\text{rep}$ of the QCL-comb, and we recorded 200 RF spectra at the output of the RF extraction channel of the QCL-comb to cover the frequency sweep. Figure~\ref{fig:capture_range} shows an example of map obtained for a NIR illumination power of 6~mW where three regimes can be observed. 

The first regime corresponds to the case where $f_\text{gen}$ is relatively far away from the natural repetition frequency of the QCL-comb, and in this case $f_\text{rep}$ stays unchanged compared to no modulation of the NIR light. 
When $f_\text{gen}$ is brought closer to $f_\text{rep}$, a pulling mechanism appears such as $f_\text{rep}$ is attracted by $f_\text{gen}$ and its frequency is hence modified compared to the case where the NIR light is not modulated.
Finally, the last regime is the injection locked regime where $f_\text{rep} = f_\text{gen}$ over the locking range $\Delta f $.
This particular evolution of $f_\text{rep}$ with respect to $f_\text{gen}$ sweeping is characteristic of Adler's formalism of coupled oscillators~\cite{Adler-ieee-1946,Siegman-lasers-1986}.

\begin{figure}[t]
\centering
\includegraphics[width=\linewidth]{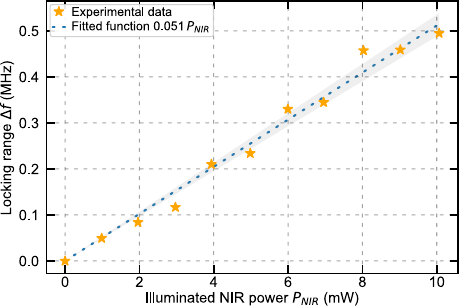}
\caption{Graph showing the locking range $\Delta f$ measured with respect to the NIR light illuminated power. Every experimental point was extracted from the recording of maps such as the one presented in Figure~\ref{fig:capture_range}. The gray shaded area represents the two standard deviation margin on the parameter estimated numerically.}
\label{fig:adler}
\end{figure}

We now take an interest of the dependency of the locking range $\Delta f$ with respect to the NIR illuminated power $ P_\text{NIR} $. For this, we record different maps such as the one presented in Figure~\ref{fig:capture_range} at different NIR powers and we extract $\Delta f$. The results are displayed in Figure~\ref{fig:adler} and a comparison of the experimental data with a fitted function shows that $\Delta f $ closely follows a linear trend with respect to $ P_\text{NIR} $.
This feature is also in accordance with Adler's formalism. As already mentioned, NIR light illumination induces a photovoltage. As we are now modulating the NIR light, a RF signal $V_\text{RF}$ is also induced in addition to the DC voltage, and $V_\text{RF}$ can be described with a similar form as the one used for $V_\text{DC}$ in equation~\eqref{eq:vinduce}. However, our experimental results shows a linear dependence between $V_\text{RF}$ and $P_\text{NIR}$, which indicates that with the power at play, the form used in equation~\eqref{eq:vinduce} can be simplified and $V_\text{RF}$ can be written as:
\begin{equation} \label{eq:RFinduce}
    V_\text{RF}(P_\text{in}) = \gamma_{0,\text{RF}} \, P_\text{in} 
\end{equation}
where $ \gamma_{0,\text{RF}} $ is the RF conversion gain at low power.
Now, using Ohm's law and equation~\eqref{eq:RFinduce}, the RF power $P_\text{RF}$ is given by:
\begin{equation}
    P_\text{RF} = \frac{1}{R_L} V_\text{RF}^2 = \frac{1}{R_L} \gamma_{0,\text{RF}}^2 P_\text{in}^2 
\end{equation}
where $R_L$ is the load resistor. $P_\text{RF} $ then follows a linear trend with respect to $ P_\text{in}^2$ and also to $ P_\text{NIR}^2$. Using Adler's law~\cite{Adler-ieee-1946,Siegman-lasers-1986}, we have in the end:
\begin{equation} \label{eq:adler}
    \Delta f = \frac{2}{Q} f_\text{rep,free} \sqrt{\frac{P_\text{RF}}{P_\text{free}}} \propto P_\text{NIR}
\end{equation}
where $f_\text{rep,free}$ is the repetition frequency of the QCL-comb in free-running, $P_\text{free}$ its power, and $Q$ the quality factor of the free-running oscillator resonator. Adler's formalism predicts that $\Delta f$ follows a square root trend with respect to $P_\text{RF}$, which in our case translates to a linear trend with respect to $P_\text{NIR}$, as can be seen with Figure~\ref{fig:adler}.

Regarding the alignment of the NIR beam, which we hereby chose so as to maximize the frequency excursion of $f_\text{rep}$ at a given NIR power and hence the voltage $V$ induced at the bounds of the QCL, we investigated the locking range in different alignment configurations. For a relevant comparison, $ P_\text{NIR} $ was kept identical, and the misalignements were chosen so as to induce a similar repetition frequency detuning. In this case, the locking range obtained between four different misalignments was found to be similar but well below the one that could be obtained by maximizing the excursion of $f_\text{rep}$.
On the contrary, for a given shift of $f_\text{rep}$ that is obtained for different values of $ P_\text{NIR} $ in different alignments of the NIR beam, we observed similar locking ranges. Hence, optimization of the locking range has to be performed in consistency with all of these aspects, i.e, a high NIR power and by maximizing the detuning excursion of $f_\text{rep}$.

\subsection{Optical spectral properties}

\begin{figure}
\centering
\includegraphics[width=\linewidth]{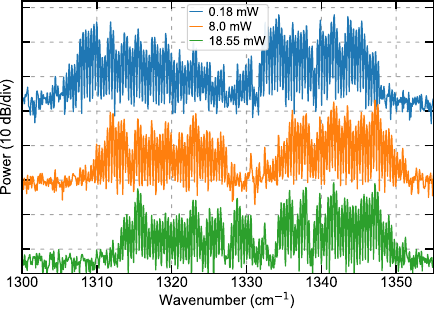}
\caption{Graph presenting several optical spectra recorded at different NIR illumination power when injection locking of the repetition frequency of the QCL-comb is performed.}
\label{fig:optical_evol}
\end{figure}

In parallel to the characterization of the repetition frequency, we also investigated the evolution of the optical spectrum delivered by the QCL with respect to $ P_\text{NIR} $ when optical RFIL is performed. Figure~\ref{fig:optical_evol} shows the spectrum retrieved using a FTIR at three different values of $ P_\text{NIR} $. Note that the spectrum at zero NIR illumination power shows no difference with the one observed at the lowest power of 0,18~mW.
One can first observe that the signal-to-noise ratio decreases with $ P_\text{NIR} $, which is linked to the evolution of the MIR power as we have seen it before with Figure~\ref{fig:param_qcl}~\textbf{(b)}.
The second observation is that the center frequency of the spectrum is slightly pushed towards higher frequencies when increasing $ P_\text{NIR} $, which is in agreement with previous work performed on CW QCLs emitting in the MIR~\cite{Chen-oe-2009,Chen-apl-2010}, or in the terahertz~\cite{alam-oe-2019a,alam-oe-2019b}. Moreover, we observed a similar behavior when operating our MIR QCL in a single operation mode which can be done by setting the pump current at a lower value.
Finally, at high illumination power, a small reduction of the bandwidth of the comb is observed with around 20 vanishing lines that are mostly located in the red part of the spectrum.

\section{Discussion and conclusion}
Stabilization of the repetition frequency of QCL-combs can be performed in various ways, and here we demonstrated a new possibility that requires no feedback loop using an intensity modulated NIR light illuminating the front facet of the QCL. We showed that with a NIR power as low as 180~µW, optical RFIL leads to a strong noise reduction of the repetition frequency. Increasing the NIR power to 5~mW allowed us to reach a residual phase noise below the symbolic value of 1~mrad without decreasing too much the MIR optical power emitted by the QCL nor by modifying too much the comb spectrum. We also observed here that the locking range of the stabilization scheme linearly follows the NIR power illuminating the front facet of the QCL, which is in good accordance with Adler's formalism.

Further investigation on this injection locking mechanism with NIR light illumination is possible, especially on the optical properties of the locked comb. For instance, SWIFT interferometry can be performed for coherence studies across the spectrum as already made for electrical RFIL~\cite{Hillbrand-natphot-2019,Schneider-lpr-2021}. One could also consider studying the comb linewidth by heterodyne beating with a fully stabilized comb~\cite{Cappelli-lpr-2016,Consolino-natcomm-2019,Chomet-optica-2024}, or a narrow-linewidth single-mode QCL~\cite{Shehzad-oe-2020}, to evidence the effect of electrical RFIL and optical RFIL on the noise correlations between $f_\text{rep}$ and $f_0$~\cite{Shehzad-oe-2020}. 
Such a study would be necessary to draw the full picture about the spectral resolution that can be achieved with a QCL-comb stabilized via optical RFIL, and hence limitations in the performance that can be obtained for MIR spectroscopy and metrological applications.
Moreover, studying the comb linewidth has the potential to provide a full comparison between optical RFIL and electrical RFIL, to assess their performance. A demonstration of full stabilization of the QCL-comb using NIR light would also be an interesting study. 
Regarding the non-resonant light used for injection locking, using a different wavelength in the NIR might offer better performance, e.g., loosen the constraint between noise reduction and MIR power loss. Moreover, the NIR beam delivery can be modified for a better coupling in the QCL and hence reducing the needed NIR power, for example by rear facet illumination using an optical fiber brought very close to the QCL~chip~\cite{alam-oe-2019b}. Also, one could consider injection at harmonics of the QCL-comb optical central frequency.

\section*{Acknowledgments}
We thank F.~Kapsalidis for processing the QCL layer grown and J.~Hillbrand for mounting the QCL-chip.
We also thank B. Chomet and C. Sirtori for sharing the data from their study~\cite{Chomet-optica-2024,Chomet-optica_sup-2024}, and for fruitful discussions.

We acknowledge funding from the Schweizerischer Nationalfonds zur Förderung der Wissenschaftlichen Forschung (Grant No. 40B2-1\_176584).

\section*{Author Declaration}
The authors have no conflicts to disclose.

\section*{Data availability}
The data that support the findings of this study are openly available in EUDAT B2SHARE at \href{http://doi.org/10.23728/b2share.e9b145dd968b4a3a957f9cdb91d84a65}{10.23728/b2share.e9b145dd968b4a3a957f9cdb91d84a65}\cite{dataset}.

\bibliography{biblio}

\end{document}